\input epsf
\newfam\scrfam
\batchmode\font\tenscr=rsfs10 \errorstopmode
\ifx\tenscr\nullfont
        \message{rsfs script font not available. Replacing with calligraphic.}
        \def\scr{\cal}
\else   
        \font\sevenscr=rsfs7
        \font\fivescr=rsfs5
        \skewchar\tenscr='177 \skewchar\sevenscr='177 \skewchar\fivescr='177
        \textfont\scrfam=\tenscr \scriptfont\scrfam=\sevenscr
        \scriptscriptfont\scrfam=\fivescr
        \def\scr{\fam\scrfam}
        \def\cal{\scr}
\fi
\catcode`\@=11
\newfam\frakfam
\batchmode\font\tenfrak=eufm10 \errorstopmode
\ifx\tenfrak\nullfont
        \message{eufm font not available. Replacing with italic.}
        
\else
	
	\font\sevenfrak=eufm7 \font\fivefrak=eufm5
        
	\textfont\frakfam=\tenfrak
	\scriptfont\frakfam=\sevenfrak \scriptscriptfont\frakfam=\fivefrak
	
\fi
\catcode`\@=\active
\newfam\msbfam
\batchmode\font\twelvemsb=msbm10 scaled\magstep1 \errorstopmode
\ifx\twelvemsb\nullfont\def\Bbb{\bf}
        
	\font\eightbbb=cmb10 at 8pt
	\message{Blackboard bold not available. Replacing with boldface.}
\else   \catcode`\@=11
        \font\tenmsb=msbm10 \font\sevenmsb=msbm7 \font\fivemsb=msbm5
        \textfont\msbfam=\tenmsb
        \scriptfont\msbfam=\sevenmsb \scriptscriptfont\msbfam=\fivemsb
        \def\Bbb{\relax\expandafter\Bbb@}
        \def\Bbb@#1{{\Bbb@@{#1}}}
        \def\Bbb@@#1{\fam\msbfam\relax#1}
        \catcode`\@=\active
	
	\font\eightbbb=msbm8
\fi
        \font\fivemi=cmmi5
        \font\sixmi=cmmi6
        \font\eightrm=cmr8              \def\xrm{\eightrm}
        \font\eightbf=cmbx8             \def\xbf{\eightbf}
        \font\eightit=cmti10 at 8pt     \def\xit{\eightit}
                
        \font\eighttt=cmtt8             
        \font\eightcp=cmcsc8
        \font\eighti=cmmi8              \def\xold{\eighti}
        \font\eightmi=cmmi8
        \font\eightib=cmmib8             \def\xbold{\eightib}
        \font\teni=cmmi10               \def\old{\teni}
        \font\tencp=cmcsc10

        \font\twelvecp=cmcsc10 scaled\magstep1
        
        \font\sixrm=cmr6
        \font\fiverm=cmr5

        \font\eightsy=cmsy8
        \font\sixsy=cmsy6
        \font\eightsl=cmsl8
        \font\sixbf=cmbx6

	 at10pt	
	\font\twelvehelvbold=phvb at12pt
	 at14pt
	\font\sixteenhelvbold=phvb at16pt

\def\noblackbox{\overfullrule=0pt}
\noblackbox

\def\eightpoint{
\def\rm{\fam0\eightrm}
\textfont0=\eightrm \scriptfont0=\sixrm \scriptscriptfont0=\fiverm
\textfont1=\eightmi  \scriptfont1=\sixmi  \scriptscriptfont1=\fivemi
\textfont2=\eightsy \scriptfont2=\sixsy \scriptscriptfont2=\fivesy
\textfont3=\tenex   \scriptfont3=\tenex \scriptscriptfont3=\tenex
\textfont\itfam=\eightit \def\it{\fam\itfam\eightit}
\textfont\slfam=\eightsl \def\sl{\fam\slfam\eightsl}
\textfont\ttfam=\eighttt \def\tt{\fam\ttfam\eighttt}
\textfont\bffam=\eightbf \scriptfont\bffam=\sixbf 
                         \scriptscriptfont\bffam=\fivebf
                         \def\bf{\fam\bffam\eightbf}
\normalbaselineskip=10pt}

\newtoks\headtext
\headline={\ifnum\pageno=1\hfill\else
	\ifodd\pageno{\eightcp\the\headtext}{ }\dotfill{ }{\old\folio}
	\else{\old\folio}{ }\dotfill{ }{\eightcp\the\headtext}\fi
	\fi}
\def\makeheadline{\vbox to 0pt{\vss\noindent\the\headline\break
\hbox to\hsize{\hfill}}
        \vskip2\baselineskip}
\newcount\infootnote
\infootnote=0
\newcount\footnotecount
\footnotecount=1
\def\foot#1{\infootnote=1
\footnote{${}^{\the\footnotecount}$}{\vtop{\baselineskip=.75\baselineskip
\advance\hsize by
-\parindent{\eightpoint\rm\hskip-\parindent
#1}\hfill\vskip\parskip}}\infootnote=0\global\advance\footnotecount by
1}
\newcount\refcount
\refcount=1
\newwrite\refwrite
\def\oldsize{\ifnum\infootnote=1\xold\else\old\fi}
\def\ref#1#2{
	\def#1{{{\oldsize\the\refcount}}\ifnum\the\refcount=1\immediate\openout\refwrite=\jobname.refs\fi\immediate\write\refwrite{\item{[{\xold\the\refcount}]} 
	#2\hfill\par\vskip-2pt}\xdef#1{{\noexpand\oldsize\the\refcount}}\global\advance\refcount by 1}
	}
\def\refout{\eightpoint\catcode`\@=11
        \xrm\immediate\closeout\refwrite
        \vskip2\baselineskip
        {\noindent\twelvecp References}\hfill\vskip\baselineskip
        \baselineskip=.75\baselineskip
        \input\jobname.refs
        \baselineskip=4\baselineskip \divide\baselineskip by 3
        \catcode`\@=\active\rm}

\def\skipref#1{\hbox to15pt{\phantom{#1}\hfill}\hskip-15pt}

\def\hepth#1{\href{http://xxx.lanl.gov/abs/hep-th/#1}{arXiv:\allowbreak
hep-th\slash{\xold#1}}}

\def\matharx#1{\href{http://xxx.lanl.gov/abs/math/#1}{arXiv:math/{\xold#1}}}
\def\arxiv#1#2{\href{http://arxiv.org/abs/#1.#2}{arXiv:\allowbreak
{\xold#1}.{\xold#2}} [hep-th]} 
\def\arxivmdg#1#2{\href{http://arxiv.org/abs/#1.#2}{arXiv:\allowbreak
{\xold#1}.{\xold#2}} [math.DG]} 
\def\jhep#1#2#3#4{\href{http://jhep.sissa.it/stdsearch?paper=#2\%28#3\%29#4}{J. High Energy Phys. {\xbold #1#2} ({\xold#3}) {\xold#4}}}

\def\CQG#1#2#3{Class. Quantum Grav. {\xbold#1} ({\xold#2}) {\xold#3}}
\def\FP#1#2#3{Fortsch. Phys. {\xbold#1} ({\xold#2}) {\xold#3}}

\def\IJMPCS#1#2#3{Int. J. Mod. Phys. Conf. Ser. {\xbf A}{\xbold#1} ({\xold#2}) {\xold#3}}

\def\JMP#1#2#3{J. Math. Phys. {\xbold#1} ({\xold#2}) {\xold#3}}
\def\JPA#1#2#3{J. Phys. {\xbf A}{\xbold#1} ({\xold#2}) {\xold#3}}
\def\LMP#1#2#3{Lett. Math. Phys. {\xbold#1} ({\xold#2}) {\xold#3}}
\def\MPLA#1#2#3{Mod. Phys. Lett. {\xbf A}{\xbold#1} ({\xold#2}) {\xold#3}}

\def\NPB#1#2#3{Nucl. Phys. {\xbf B}{\xbold#1} ({\xold#2}) {\xold#3}}

\def\PLB#1#2#3{Phys. Lett. {\xbf B}{\xbold#1} ({\xold#2}) {\xold#3}}
\def\PR#1#2#3{Phys. Rept. {\xbold#1} ({\xold#2}) {\xold#3}}
\def\PRD#1#2#3{Phys. Rev. {\xbf D}{\xbold#1} ({\xold#2}) {\xold#3}}
\def\PRL#1#2#3{Phys. Rev. Lett. {\xbold#1} ({\xold#2}) {\xold#3}}

\def\TMP#1#2#3{Theor. Math. Phys. {\xbold#1} ({\xold#2}) {\xold#3}}
\def\TMF#1#2#3{Teor. Mat. Fiz. {\xbold#1} ({\xold#2}) {\xold#3}}
\newcount\sectioncount
\sectioncount=0
\def\section#1#2{\global\eqcount=0
	\global\subsectioncount=0
        \global\advance\sectioncount by 1
	\ifnum\sectioncount>1
	        \vskip2\baselineskip
	\fi
\line{\twelvecp\the\sectioncount. #2\hfill}
       \vskip.5\baselineskip\noindent
        \xdef#1{{\old\the\sectioncount}}}
\newcount\subsectioncount
\def\subsection#1#2{\global\advance\subsectioncount by 1
\vskip.75\baselineskip\noindent\line{\tencp\the\sectioncount.\the\subsectioncount. #2\hfill}\nobreak\vskip.4\baselineskip\nobreak\noindent\xdef#1{{\old\the\sectioncount}.{\old\the\subsectioncount}}}
\def\immediatesubsection#1#2{\global\advance\subsectioncount by 1
\vskip-\baselineskip\noindent
\line{\tencp\the\sectioncount.\the\subsectioncount. #2\hfill}
	\vskip.5\baselineskip\noindent
	\xdef#1{{\old\the\sectioncount}.{\old\the\subsectioncount}}}
\newcount\subsubsectioncount
\def\subsubsection#1#2{\global\advance\subsubsectioncount by 1
\vskip.75\baselineskip\noindent\line{\tencp\the\sectioncount.\the\subsectioncount.\the\subsubsectioncount. #2\hfill}\nobreak\vskip.4\baselineskip\nobreak\noindent\xdef#1{{\old\the\sectioncount}.{\old\the\subsectioncount}.{\old\the\subsubsectioncount}}}
\newcount\appendixcount
\appendixcount=0
\def\appendix#1{\global\eqcount=0
        \global\advance\appendixcount by 1
        \vskip2\baselineskip\noindent
        \ifnum\the\appendixcount=1
        \hbox{\twelvecp Appendix A: #1\hfill}\vskip\baselineskip\noindent\fi
    \ifnum\the\appendixcount=2
        \hbox{\twelvecp Appendix B: #1\hfill}\vskip\baselineskip\noindent\fi
    \ifnum\the\appendixcount=3
        \hbox{\twelvecp Appendix C: #1\hfill}\vskip\baselineskip\noindent\fi}
\def\acknowledgements{\vskip2\baselineskip\noindent
        \underbar{\it Acknowledgements:}\ }
\newcount\eqcount
\eqcount=0
\def\Eqn#1{\global\advance\eqcount by 1
\ifnum\the\sectioncount=0
	\xdef#1{{\noexpand\oldsize\the\eqcount}}
	\eqno({\oldstyle\the\eqcount})
\else
        \ifnum\the\appendixcount=0
\xdef#1{{\noexpand\oldsize\the\sectioncount}.{\noexpand\oldsize\the\eqcount}}
                \eqno({\oldstyle\the\sectioncount}.{\oldstyle\the\eqcount})\fi
        \ifnum\the\appendixcount=1
	        \xdef#1{{\noexpand\oldstyle A}.{\noexpand\oldstyle\the\eqcount}}
                \eqno({\oldstyle A}.{\oldstyle\the\eqcount})\fi
        \ifnum\the\appendixcount=2
	        \xdef#1{{\noexpand\oldstyle B}.{\noexpand\oldstyle\the\eqcount}}
                \eqno({\oldstyle B}.{\oldstyle\the\eqcount})\fi
        \ifnum\the\appendixcount=3
	        \xdef#1{{\noexpand\oldstyle C}.{\noexpand\oldstyle\the\eqcount}}
                \eqno({\oldstyle C}.{\oldstyle\the\eqcount})\fi
\fi}
\def\eqn{\global\advance\eqcount by 1
\ifnum\the\sectioncount=0
	\eqno({\oldstyle\the\eqcount})
\else
        \ifnum\the\appendixcount=0
                \eqno({\oldstyle\the\sectioncount}.{\oldstyle\the\eqcount})\fi
        \ifnum\the\appendixcount=1
                \eqno({\oldstyle A}.{\oldstyle\the\eqcount})\fi
        \ifnum\the\appendixcount=2
                \eqno({\oldstyle B}.{\oldstyle\the\eqcount})\fi
        \ifnum\the\appendixcount=3
                \eqno({\oldstyle C}.{\oldstyle\the\eqcount})\fi
\fi}
\def\multi{\global\advance\eqcount by 1}
\def\multieqn#1{({\oldstyle\the\sectioncount}.{\oldstyle\the\eqcount}\hbox{#1})}
\def\multiEqn#1#2{\xdef#1{{\oldstyle\the\sectioncount}.{\old\the\eqcount}#2}
        ({\oldstyle\the\sectioncount}.{\oldstyle\the\eqcount}\hbox{#2})}
\def\multiEqnAll#1{\xdef#1{{\oldstyle\the\sectioncount}.{\old\the\eqcount}}}
\newcount\tablecount
\tablecount=0
\def\Table#1#2{\global\advance\tablecount by 1
       \xdef#1{\the\tablecount}
       \vskip2\parskip
       \centerline{\it Table \the\tablecount: #2}
       \vskip2\parskip}
\newtoks\url
\def\Href#1#2{\catcode`\#=12\url={#1}\catcode`\#=\active#2}
\def\href#1#2{{#2}}

\parskip=3.5pt plus .3pt minus .3pt
\baselineskip=14pt plus .1pt minus .05pt
\lineskip=.5pt plus .05pt minus .05pt
\lineskiplimit=.5pt
\abovedisplayskip=18pt plus 4pt minus 2pt
\belowdisplayskip=\abovedisplayskip
\hsize=14cm
\vsize=19cm
\hoffset=1.5cm
\voffset=1.8cm
\frenchspacing
\footline={}
\raggedbottom

\newskip\origparindent
\origparindent=\parindent

\def\*{\partial}
\def\punkt{\,\,.}
\def\komma{\,\,,}

\def\={\!=\!}
\def\small#1{{\hbox{$#1$}}}

\def\fraction#1{\small{1\over#1}}
\def\fr{\fraction}
\def\Fraction#1#2{\small{#1\over#2}}
\def\Fr{\Fraction}

\def\eg{{\it e.g.}}

\def\ie{{\it i.e.}}

\def\nlni{\hfill\break}

\def\a{\alpha}
\def\b{\beta}

\def\g{\gamma}
\def\l{\lambda}

\def\RR{{\Bbb R}}




\def\textfrac#1#2{\raise .45ex\hbox{\the\scriptfont0 #1}\nobreak\hskip-1pt/\hskip-1pt\hbox{\the\scriptfont0 #2}}

\def\LL{{\cal L}}
\def\leftbr{[\![}
\def\rightbr{]\!]}


\def\frac{\Fr}

\def\mathbb{\Bbb}



\def\LL{{\cal L}}
\def\leftbr{[\![}
\def\rightbr{]\!]}

\def\ti#1{(#1^{-1})^t}

\def\LL{{\cal L}}
\def\leftbr{[\![}
\def\rightbr{]\!]}

\def\vvecc#1{\vec{\hskip1pt #1}}

\def\vxi{\vvecc\xi}
\def\veta{\vvecc\eta}
\def\vzeta{\vvecc\zeta}

\def\ad{\hbox{ad}}
\def\adxi{\ad_\xi}

\ref\HohmZwiebachLarge{O. Hohm and B. Zwiebach, {\xit ``Large gauge
transformations in double field theory''}, \jhep{13}{02}{2013}{075}
[\arxiv{1207}{4198}].} 

\ref\AschieriEtAl{P. Aschieri, I. Bakovi\'c, B. Jur\v co and
P. Schupp, {\xit ``Noncommutative gerbes and deformation
quantization''}, \hepth{0206101}.}

\ref\BermanCederwallKleinschmidtThompson{D.S. Berman, M. Cederwall,
A. Kleinschmidt and D.C. Thompson, {\xit ``The gauge structure of
generalised diffeomorphisms''}, \jhep{13}{01}{2013}{64} [\arxiv{1208}{5884}].}

\ref\CederwallUfoldbranes{M. Cederwall, {\xit ``M-branes on U-folds''},
in proceedings of 7th International Workshop ``Supersymmetries and
Quantum Symmetries'' Dubna, 2007 [\arxiv{0712}{4287}].}

\ref\BermanPerryGen{D.S. Berman and M.J. Perry, {\xit ``Generalised
geometry and M-theory''}, \jhep{11}{06}{2011}{074} [\arxiv{1008}{1763}].}    

\ref\BermanMusaevThompson{D.S. Berman, E.T. Musaev and D.C. Thompson,
{\xit ``Duality invariant M-theory: gaugings via Scherk--Schwarz
reduction}, \jhep{12}{10}{2012}{174} [\arxiv{1208}{0020}].}

\ref\UdualityMembranes{V. Bengtsson, M. Cederwall, H. Larsson and
B.E.W. Nilsson, {\xit ``U-duality covariant
membranes''}, \jhep{05}{02}{2005}{020} [\hepth{0406223}].}

\ref\ObersPiolineU{N.A. Obers and B. Pioline, {\xit ``U-duality and M-theory''},
\PR{318}{1999}{113}, 
\nlni [\hepth{9809039}].}

\ref\BermanGodazgarPerry{D.S. Berman, H. Godazgar and M.J. Perry,
{\xit ``SO(5,5) duality in M-theory and generalized geometry''},
\PLB{700}{2011}{65} [\arxiv{1103}{5733}].} 

\ref\BermanMusaevPerry{D.S. Berman, E.T. Musaev and M.J. Perry,
{\xit ``Boundary terms in generalized geometry and doubled field theory''},
\PLB{706}{2011}{228} [\arxiv{1110}{97}].} 

\ref\BermanGodazgarGodazgarPerry{D.S. Berman, H. Godazgar, M. Godazgar  
and M.J. Perry,
{\xit ``The local symmetries of M-theory and their formulation in
generalised geometry''}, \jhep{12}{01}{2012}{012}
[\arxiv{1110}{3930}].} 

\ref\BermanGodazgarPerryWest{D.S. Berman, H. Godazgar, M.J. Perry and
P. West,
{\xit ``Duality invariant actions and generalised geometry''}, 
\jhep{12}{02}{2012}{108} [\arxiv{1111}{0459}].} 

\ref\CoimbraStricklandWaldram{A. Coimbra, C. Strickland-Constable and
D. Waldram, {\xit ``$E_{d(d)}\times\hbox{\eightbbb R}^+$ generalised geometry,
connections and M theory'' }, \arxiv{1112}{3989}.} 

\ref\CremmerPopeI{E. Cremmer, B. Julia, H. L\"u and C.N. Pope,
{\xit ``Dualisation of dualities. I.''}, \NPB{523}{1998}{73} [\hepth{9710119}].}

\ref\HullT{C.M. Hull, {\xit ``A geometry for non-geometric string
backgrounds''}, \jhep{05}{10}{2005}{065} [\hepth{0406102}].}

\ref\HullM{C.M. Hull, {\xit ``Generalised geometry for M-theory''},
\jhep{07}{07}{2007}{079} [\hepth{0701203}].}

\ref\HullDoubled{C.M. Hull, {\xit ``Doubled geometry and
T-folds''}, \jhep{07}{07}{2007}{080}
[\hepth{0605149}].}

\ref\HullTownsend{C.M. Hull and P.K. Townsend, {\xit ``Unity of
superstring dualities''}, \NPB{438}{1995}{109} [\hepth{9410167}].}

\ref\PalmkvistHierarchy{J. Palmkvist, {\xit ``Tensor hierarchies,
Borcherds algebras and $E_{11}$''}, \jhep{12}{02}{2012}{066}
[\arxiv{1110}{4892}].} 

\ref\deWitNicolaiSamtleben{B. de Wit, H. Nicolai and H. Samtleben,
{\xit ``Gauged supergravities, tensor hierarchies, and M-theory''},
\jhep{02}{08}{2008}{044} [\arxiv{0801}{1294}].}

\ref\deWitSamtleben{B. de Wit and H. Samtleben,
{\xit ``The end of the $p$-form hierarchy''},
\jhep{08}{08}{2008}{015} [\arxiv{0805}{4767}].}

\ref\CederwallJordanMech{M.~Cederwall, {\xit ``Jordan algebra
dynamics''}, \PLB{210}{1988}{169}.} 

\ref\BerkovitsNekrasovCharacter{N. Berkovits and N. Nekrasov, {\xit
    ``The character of pure spinors''}, \LMP{74}{2005}{75}
  [\hepth{0503075}].}

\ref\HitchinLectures{N. Hitchin, {``\xit Lectures on generalized
geometry''}, \arxiv{1010}{2526}.}

\ref\KoepsellNicolaiSamtleben{K. Koepsell, H. Nicolai and
H. Samtleben, {\xit ``On the Yangian $[Y(e_8)]$ quantum symmetry of
maximal supergravity in two dimensions''}, \jhep{99}{04}{1999}{023}
[\hepth{9903111}].}

\ref\HohmHullZwiebachI{O. Hohm, C.M. Hull and B. Zwiebach, {\xit ``Background
independent action for double field
theory''}, \jhep{10}{07}{2010}{016} [\arxiv{1003}{5027}].}

\ref\HohmHullZwiebachII{O. Hohm, C.M. Hull and B. Zwiebach, {\xit
``Generalized metric formulation of double field theory''},
\jhep{10}{08}{2010}{008} [\arxiv{1006}{4823}].} 

\ref\HohmZwiebach{O. Hohm and B. Zwiebach, {\xit ``On the Riemann
tensor in double field theory''}, \jhep{12}{05}{2012}{126}
[\arxiv{1112}{5296}].} 

\ref\WestEEleven{P. West, {\xit ``$E_{11}$ and M theory''},
\CQG{18}{2001}{4443} [\hepth{0104081}].}

\ref\AndriotLarforsLustPatalong{D. Andriot, M. Larfors, D. L\"ust and
P. Patalong, {\xit ``A ten-dimensional action for non-geometric
fluxes''}, \jhep{11}{09}{2011}{134} [\arxiv{1106}{4015}].}

\ref\AndriotHohmLarforsLustPatalongI{D. Andriot, O. Hohm, M. Larfors,
D. L\"ust and 
P. Patalong, {\xit ``A geometric action for non-geometric
fluxes''}, \PRL{108}{2012}{261602} [\arxiv{1202}{3060}].}

\ref\AndriotHohmLarforsLustPatalongII{D. Andriot, O. Hohm, M. Larfors,
D. L\"ust and 
P. Patalong, {\xit ``Non-geometric fluxes in supergravity and double
field theory''}, \FP{60}{2012}{1150} [\arxiv{1204}{1979}].}

\ref\DamourHenneauxNicolai{T. Damour, M. Henneaux and H. Nicolai,
{\xit ``Cosmological billiards''}, \CQG{20}{2003}{R145} [\hepth{0212256}].}

\ref\DamourNicolai{T. Damour and H. Nicolai, 
{\xit ``Symmetries, singularities and the de-emergence of space''},
\arxiv{0705}{2643}.}

\ref\EHTP{F. Englert, L. Houart, A. Taormina and P. West,
{\xit ``The symmetry of M theories''},
\jhep{03}{09}{2003}{020}2003 [\hepth{0304206}].}

\ref\PachecoWaldram{P.P. Pacheco and D. Waldram, {\xit ``M-theory,
exceptional generalised geometry and superpotentials''},
\jhep{08}{09}{2008}{123} [\arxiv{0804}{1362}].}

\ref\DamourHenneauxNicolaiII{T. Damour, M. Henneaux and H. Nicolai,
{\xit ``$E_{10}$ and a 'small tension expansion' of M theory''},
\PRL{89}{2002}{221601} [\hepth{0207267}].}

\ref\KleinschmidtNicolai{A. Kleinschmidt and H. Nicolai, {\xit
``$E_{10}$ and $SO(9,9)$ invariant supergravity''},
\jhep{04}{07}{2004}{041} [\hepth{0407101}].}

\ref\WestII{P.C. West, {\xit ``$E_{11}$, $SL(32)$ and central charges''},
\PLB{575}{2003}{333} [\hepth{0307098}].}

\ref\KleinschmidtWest{A. Kleinschmidt and P.C. West, {\xit
``Representations of $G^{+++}$ and the r\^ole of space-time''},
\jhep{04}{02}{2004}{033} [\hepth{0312247}].}

\ref\WestIII{P.C. West, {\xit ``$E_{11}$ origin of brane charges and
U-duality multiplets''}, \jhep{04}{08}{2004}{052} [\hepth{0406150}].}

\ref\PiolineWaldron{B. Pioline and A. Waldron, {\xit ``The automorphic
membrane''}, \jhep{04}{06}{2004}{009} [\hepth{0404018}].}

\ref\WestBPS{P.C. West, {\xit ``Generalised BPS conditions''},
\arxiv{1208}{3397}.}

\ref\PalmkvistBorcherds{J. Palmkvist, {\xit ``Borcherds and Kac--Moody
  extensions of simple finite-dimensional Lie algebras''}, \arxiv{1203}{5107}.}

\ref\ParkSuh{J.-H. Park and Y. Suh, {\xit ``U-geometry: SL(5)''},
\arxiv{1302}{1652}.} 

\ref\CederwallMinimalExcMult{M. Cederwall, {\xit ``Non-gravitational 
exceptional supermultiplets''}, 
\jhep{13}{07}{2013}{025}
[\arxiv{1302}{6737}].} 

\ref\PalmkvistDual{J. Palmkvist, {\xit ``The tensor hierarchy
algebra''}, \arxiv{1305}{0018}.}

\ref\CederwallPalmkvistSerre{M. Cederwall and J. Palmkvist, {\xit
    ``Serre relations, constraints and partition functions''}, to appear.}

\ref\CoimbraStricklandWaldramII{A. Coimbra, C. Strickland-Constable and
D. Waldram, {\xit ``Supergravity as generalised geometry II:
$E_{d(d)}\times\hbox{\eightbbb R}^+$ and M theory''}, \arxiv{1212}{1586}.}  

\ref\HohmZwiebachGeometry{O. Hohm and B. Zwiebach, {\xit ``Towards an
invariant geometry of double field theory''}, \arxiv{1212}{1736}.} 

\ref\JeonLeeParkRR{I. Jeon, K. Lee and J.-H. Park, {\xit
``Ramond--Ramond cohomology and O(D,D) T-duality''},
\jhep{12}{09}{2012}{079} [\arxiv{1206}{3478}].} 

\ref\PureSGI{M. Cederwall, {\xit ``Towards a manifestly supersymmetric
    action for D=11 supergravity''}, \jhep{10}{01}{2010}{117}
    [\arxiv{0912}{1814}].}  

\ref\PureSGII{M. Cederwall, 
{\xit ``D=11 supergravity with manifest supersymmetry''},
    \MPLA{25}{2010}{3201} [\arxiv{1001}{0112}].}

\ref\CremmerLuPopeStelle{E. Cremmer, H. L\"u, C.N. Pope and
K.S. Stelle, {\xit ``Spectrum-generating symmetries for BPS solitons''},
\NPB{520}{1998}{132} [\hepth{9707207}].}

\ref\JeonLeeParkI{I. Jeon, K. Lee and J.-H. Park, {\xit ``Differential
geometry with a projection: Application to double field theory''},
\jhep{11}{04}{2011}{014} [\arxiv{1011}{1324}].}

\ref\JeonLeeParkII{I. Jeon, K. Lee and J.-H. Park, {\xit ``Stringy
differential geometry, beyond Riemann''}, 
\PRD{84}{2011}{044022} [\arxiv{1105}{6294}].}

\ref\JeonLeeParkIII{I. Jeon, K. Lee and J.-H. Park, {\xit
``Supersymmetric double field theory: stringy reformulation of supergravity''},
\PRD{85}{2012}{081501} [\arxiv{1112}{0069}].}

\ref\HohmKwak{O. Hohm and S.K. Kwak, {\xit ``$N=1$ supersymmetric
double field theory''}, \jhep{12}{03}{2012}{080} [\arxiv{1111}{7293}].}

\ref\HohmKwakFrame{O. Hohm and S.K. Kwak, {\xit ``Frame-like geometry
of double field theory''}, \JPA{44}{2011}{085404} [\arxiv{1011}{4101}].}

\ref\HohmKwakZwiebachI{O. Hohm, S.K. Kwak and B. Zwiebach, {\xit
``Unification of type II strings and T-duality''},
\PRL{107}{2011}{171603} [\arxiv{1106}{5452}].}

\ref\HohmKwakZwiebachII{O. Hohm, S.K. Kwak and B. Zwiebach, {\xit
``Double field theory of type II strings''}, \jhep{11}{09}{2011}{013}
[\arxiv{1107}{0008}].} 

\ref\Hillmann{C. Hillmann, {\xit ``Generalized $E_{7(7)}$ coset
dynamics and $D=11$ supergravity''}, \jhep{09}{03}{2009}{135}
[\arxiv{0901}{1581}].}

\ref\AldazabalGranaMarquesRosabal{G. Aldazabal, M. Gra\~na,
D. Marqu\'es and J.A. Rosabal, {\xit ``Extended geometry and gauged
maximal supergravity''}, \arxiv{1302}{5419}.}

\ref\DuffDualityString{M.J. Duff, {\xit ``Duality rotations in string
theory''}, \NPB{335}{1990}{610}.}

\ref\DuffDualityMembrane{M.J. Duff and J.X. Lu, {\xit ``Duality
rotations in membrane theory''}, \NPB{347}{1990}{394}.}

\ref\HohmLustZwiebach{O. Hohm, D. L\"ust and B. Zwiebach, {\xit ``The
spacetime of double field theory: Review, remarks and outlook''},
\arxiv{1309}{2977}.} 

\ref\HullEtAlGerbes{C.M. Hull, U. Lindstr\"om, M. Ro\v cek, R. von
Unge and M. Zabzine, {\xit ``Generalized K\"ahler geometry and
gerbes''}, \jhep{09}{10}{2009}{062} [\arxiv{0811}{3615}].}

\ref\Kikuchi{T.~Kikuchi, T.~Okada and Y.~Sakatani,
  {\xit ``Rotating string in doubled geometry with generalized isometries''},
  \PRD{86}{2012}{046001}
  [\arxiv{1205}{5549}].}

\ref\KikuchiGL{  T.~Kikuchi, T.~Okada and Y.~Sakatani,
  {\xit ``Generalized Lie transported string in T-fold with
  generalized Killing vector''}, 
  \IJMPCS{21}{2013}{169}.}

\ref\Banks{T.~Banks and L.J.~Dixon,
  {\xit ``Constraints on string vacua with space-time supersymmetry''},
  \NPB{307}{1988}{93}.}
 
\ref\Park{J.-H.~Park,
  {\xit ``Comments on double field theory and diffeomorphisms''},
  \jhep{13}{06}{2013}{098}
  [\arxiv{1304}{5946}].}

\ref\ParkLeeGauge{K.~Lee and J.-H.~Park,
  {\xit ``Covariant action for a string in doubled yet gauged spacetime''}
  \arxiv{1307}{8377}.}

\ref\Borisov{A.B.~Borisov and V.I.~Ogievetsky,
  {\xit ``Theory of dynamical affine and conformal symmetries as
  gravity theory''}, 
  \TMP{21}{1975}{1179}
   [\TMF{21}{1974}{329}].}

\ref\Tseytlin{A.A.~Tseytlin,
  {\xit ``Duality symmetric closed string theory and interacting
  chiral scalars''}, 
  \NPB{350}{1991}{395}.}


\ref\Shigemori{J.~de Boer and M.~Shigemori,
  {\xit ``Exotic branes and non-geometric backgrounds''},
  \PRL{104}{2010}{251603}
  [\arxiv{1004}{2521}].}

\ref\ShigemoriEB{J.~de Boer and M.~Shigemori,
  {\xit ``Exotic branes in string theory''},
  \PR{532}{2013}{65}
  [\arxiv{1209}{6056}].}

\ref\Hitchin{N.J.~Hitchin,
 {\xit ``Lectures on special Lagrangian submanifolds''},
  \matharx/9907034.}

\ref\LustNA{R.~Blumenhagen, M.~Fuchs, D.~L\"ust and R.~Sun,
 {\xit ``Non-associative deformations of geometry in double field theory''},
  \arxiv{1312}{0719}.}

\ref\Blumenhagen{R.~Blumenhagen,
  {\xit ``Nonassociativity in string theory''},
  \arxiv{1112}{4611}.}

\ref\BlumenhagenAlgebroid{R. Blumenhagen, A. Deser, E. Plauschinn and
F. Rennecke, {\xit ``Non-geometric strings, symplectic gravity and
differential geometry of Lie algebroids ''}, 
\jhep{13}{02}{2013}{122} [\arxiv{1304}{2784}].}

\ref\BlumenhagenAlgebroidII{R. Blumenhagen, A. Deser, E. Plauschinn,
F. Rennecke and C. Schmid, {\xit ``The intriguing structure of
non-geometric frames in string theory''}, \arxiv{1304}{2784}.}

\ref\Bakas{I.~Bakas and D.~L\"ust,
  {\xit ``3-cocycles, non-associative star-products and the magnetic
  paradigm of R-flux string vacua''}, 
  \arxiv{1309}{3172}.}

\ref\Marques{G.~Aldazabal, D.~Marqu\'es and C.~Nu\~nez,
  {\xit ``Double field theory: A pedagogical review''},
  \CQG{30}{2013}{163001}
  [\arxiv{1305}{1907}].}

\ref\BermanThompson{D.S.~Berman and D.C.~Thompson,
  {\xit``Duality symmetric string and M-theory''},
  \arxiv{1306}{2643}.}

\ref\SiegelI{W.~Siegel,
  {\xit ``Two vierbein formalism for string inspired axionic gravity''},
  \PRD{47}{1993}{5453}
  [\hepth{9302036}].}

\ref\SiegelII{ W.~Siegel,
  {\xit ``Superspace duality in low-energy superstrings''},
  \PRD{48}{1993}{2826}
  [\hepth{9305073}]}.

\ref\SiegelIII{W.~Siegel,
  {\xit ``Manifest duality in low-energy superstrings''},
  in Berkeley 1993, Proceedings, Strings '93 353
  [\hepth{9308133}].}

\ref\Belov{D.M.~Belov, C.M.~Hull and R.~Minasian,
  {\xit ``T-duality, gerbes and loop spaces''},
  \arxiv{0710}{5151}.}

\ref\HullZwiebachCourant{C.~Hull and B.~Zwiebach,
  {\xit ``The gauge algebra of double field theory and Courant brackets''},
  \jhep{09}{09}{2009}{090}
  [\arxiv{0908}{1792}].}

\ref\Sugawara{S.~Kawai and Y.~Sugawara,
  {\xit ``Mirrorfolds with K3 fibrations''},
  \jhep{08}{02}{2008}{065}
  [\arxiv{0711}{1045}].}



\ref\Szabo{D. Mylonas, P. Schupp and R.J. Szabo,
 {\xit ``Membrane sigma-models and quantization of non-geometric flux
backgrounds"}
  \jhep{12}{09}{2012}{012}
  [\arxiv{1207}{0926}].}

\ref\PachecoWaldram{P.~P.~Pacheco and D.~Waldram,
  {\xit ``M-theory, exceptional generalised geometry and superpotentials''},
  \jhep{08}{09}{2008}{123}
  [\arxiv{0804}{1362}].}

\ref\CederwallI{M.~Cederwall, J.~Edlund and A.~Karlsson,
  {\xit ``Exceptional geometry and tensor fields''},
  \jhep{13}{07}{2013}{028}
  [\arxiv{1302}{6736}].}

\ref\CederwallII{ M.~Cederwall,
  {\xit ``Non-gravitational exceptional supermultiplets''},
  \jhep{13}{07}{2013}{025}
  [\arxiv{1302}{6737}].}

\ref\SambtlebenHohmI{O.~Hohm and H.~Samtleben,
  {\xit ``Exceptional field theory I: $E_{6(6)}$ covariant form of
  M-theory and type IIB''}, 
  \arxiv{1312}{0614}.}

\ref\SambtlebenHohmII{O.~Hohm and H.~Samtleben,
  {\xit ``Exceptional field theory II: $E_{7(7)}$''},
  \arxiv{1312}{4542}.}

\ref\Diego{G.~Aldazabal, M.~Gra\~na, D.~Marqu\'es and J.A.~Rosabal,
  {\xit ``The gauge structure of exceptional field theories and the
  tensor hierarchy''}, 
  \arxiv{1312}{4549}.}

\ref\Riccioni{F.~Riccioni and P.~C.~West,
  {\xit``E(11)-extended spacetime and gauged supergravities''},
  \jhep{08}{02}{2008}{039}
  [\arxiv{0712}{1795}].}

\ref\Rocen{A.~Roc\'en and P.~West,
  {\xit``E11, generalised space-time and IIA string theory: the R-R sector''},
  \arxiv{1012}{2744}.}

\ref\Duff{M.~J.~Duff,
  {\xit ``Duality rotations in string theory''},
  \NPB{335}{1990}{610}.}

\ref\VaismanI{I.~Vaisman,
  {\xit ``On the geometry of double field theory''},
  \JMP{53}{2012}{033509}
  [\arxivmdg{1203}{0836}].}

\ref\VaismanII{I.~Vaisman,
  {\xit ``Towards a double field theory on para-Hermitian manifolds''},
  \arxivmdg{1209}{0152}.}

\ref\VaismanIII{I.~Vaisman,
  {\xit ``Geometry on big-tangent manifolds''},
  \arxivmdg{1303}{0658}.}

\ref\BlairMalekPark{C.D.A. Blair, E. Malek and J.-H. Park,
{\xit ``M-theory and F-theory from a duality manifest action''},
\arxiv{1311}{5109}.}

\ref\StricklandConstableSubsectors{C. Strickland-Constable, 
{\xit ``Subsectors, Dynkin diagrams and new generalised geometries''},
\arxiv{1310}{4196}.}

\ref\BlairMalekRouth{C.D.A. Blair, E. Malek and A.J. Routh,
{\xit ``An O(D,D) invariant Hamiltonian action for the superstring''},
\arxiv{1308}{4829}.}

\ref\FreidelLeighMinic{L. Freidel, R.G. Leigh and Dj. Mini\'c,
{\xit ``Born reciprocity in string theory and the nature of spacetime''},
\arxiv{1307}{7080}.}

\ref\BermanBlairMalekPerryODD{D.S. Berman, C.D.A. Blair, E. Malek and
M.J. Perry, 
{\xit ``The $O_{D,D}$ geometry of string theory''},
\arxiv{1303}{6727}.} 




\ref\HohmSamtleben{O. Hohm and H. Samtleben, {\xit ``Exceptional form
of $D=11$ supergravity''}, \arxiv{1308}{1673}.}

\ref\HullZwiebachDFT{C.~Hull and B.~Zwiebach, {\xit ``Double Field Theory''}
  JHEP {\bf 0909} (2009) 099
  [arXiv:0904.4664 [hep-th]].}

\ref\Papadopoulos{  G.~Papadopoulos,
  {\xit ``Seeking the balance: Patching double and exceptional field theories''}
  arXiv:1402.2586 [hep-th].}

\ref\Hullmxa{ C.~ M.~Hull,
  {\xit``Finite Gauge Transformations and Geometry in Double Field Theory''},
  arXiv:1406.7794 [hep-th].}


\null\vskip-1.5cm

\line{
\epsfysize=9mm
\epsffile{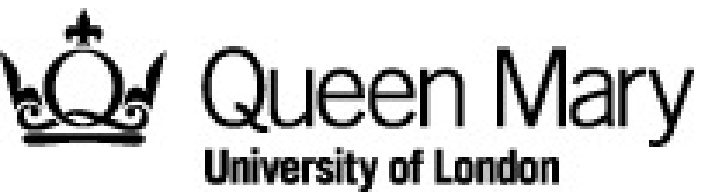}
\hskip3mm\epsfysize=11mm
\epsffile{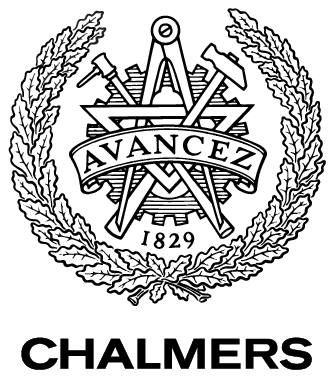}
\hskip3mm\epsfysize=9mm
\epsffile{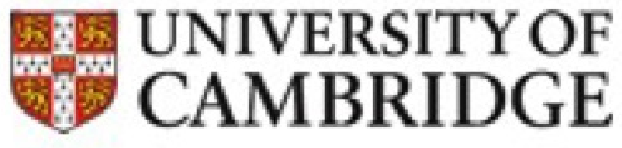}
\hfill}
\vskip-13mm
\vtop{\baselineskip=.75\baselineskip
\line{\hfill\xrm QMUL-PH-{\old13}-{\old14}}
\line{\hfill\xrm Gothenburg preprint}
\line{\hfill\xrm January, {\xold2014}}
}

\vskip\parskip
\line{\hrulefill}

\headtext={Berman, Cederwall, Perry: 
``Global aspects of double geometry''}

\vfill
\vskip.5cm

\centerline{\sixteenhelvbold
Global aspects of double geometry}

\vfill
\vskip.3cm

\centerline{\twelvehelvbold David S. Berman}

\vskip\parskip

\centerline{\it School of Physics and Astronomy} 
\centerline{\it Queen Mary University of London}
\centerline{\it Mile End Road} 
\centerline{\it London E1 4NS, England}

\vfill

\centerline{\twelvehelvbold Martin Cederwall}

\vskip\parskip

\centerline{\it Dept. of Fundamental Physics}
\centerline{\it Chalmers University of Technology}
\centerline{\it SE 412 96 Gothenburg, Sweden}

\vfill

\centerline{\twelvehelvbold Malcolm J. Perry}

\vskip\parskip

\centerline{\it DAMTP}
\centerline{\it Centre for Mathematical Sciences}
\centerline{\it Wilberforce Road}
\centerline{\it Cambridge CB3 0WA, England}

\vfill
\vskip.3cm

{\narrower\noindent \underbar{Abstract:} 
We consider the concept of a generalised manifold in the $O(d,d)$
setting, \ie, in double geometry. The conjecture by Hohm and Zwiebach
for the form of finite generalised diffeomorphisms is shown to hold.
Transition functions on overlaps are defined. Triple overlaps are
trivial concerning their action on coordinates, but non-trivial on
fields, including the generalised metric. A generalised manifold is an
ordinary manifold, but the generalised metric on the manifold carries a gerbe
structure. We show how the abelian behaviour of the gerbe is embedded
in the non-abelian T-duality group.
We also comment on possibilities and difficulties in the U-duality
setting.
\smallskip}
\vfill

\font\xxtt=cmtt6

\vtop{\baselineskip=.6\baselineskip\xxtt
\line{\hrulefill}
\catcode`\@=11
\line{email: d.s.berman@qmul.ac.uk, martin.cederwall@chalmers.se, 
malcolm@damtp.cam.ac.uk\hfill}
\catcode`\@=\active
}

\eject

\def\textfrac#1#2{\raise .45ex\hbox{\the\scriptfont0 #1}\nobreak\hskip-1pt/\hskip-1pt\hbox{\the\scriptfont0 #2}}

\section\IntroSect{Introduction}Double field theory is by now becoming
a well developed subject. After the interesting early work of [\Duff],
[\Tseytlin] constructed a version of the string in a 
doubled space. Siegel then showed that there was a new type of
geometry which may be used to describe supergravity in a duality
covariant way [\SiegelI\skipref\SiegelII-\SiegelIII]. [\HullT] showed
that T-folds have a natural formulation as a geometric doubled
space. It also introduced a doubled sigma model and applied it to such
cases. [\HullDoubled] developed this further and introduced a lot of
the ideas behind what would become double field theory. Then in 2009
double field theory (DFT) was introduced with the seminal work of Hull
and Zwiebach [\HullZwiebachDFT]. DFT is much more than duality
covariant versions of supergravity since it allows dynamics in all
doubled dimensions. 
This was then followed up and developed in various directions in works
such as
[\HohmHullZwiebachI\skipref\HohmHullZwiebachII\skipref\HohmKwakFrame-\HohmKwak]
and also with the work of Park and collaborators
[\JeonLeeParkI\skipref\JeonLeeParkII-\JeonLeeParkIII]. For recent
reviews on 
this subject see refs. [\Marques\skipref\BermanThompson-\HohmLustZwiebach]. 

In double field theory and in the duality covariant forms of
supergravity, the metric may be 
constructed through the non-linear realisation [\Borisov] of the
$O(d,d)/O(d) \times O(d)$ coset. This same construction is central  to
the $E_{11}$ programme of West and others
[\WestEEleven\skipref\EHTP\skipref\WestII\skipref\KleinschmidtWest\skipref\WestIII-\WestBPS]. As
such the duality manifest form of the Bosonic sector of supergravity
is somewhat of a truncation of the $E_{11}$ theory, 
a fact used to construct the RR sector of double field theory in
ref. [\Rocen,\HohmKwakZwiebachI]. 
Of course, the $E_{11}$ programme is much more
ambitious and also provides many other important results such as
amongst others an explanation of all the gauged maximal
supergravities in five dimensions [\Riccioni].  

There have been many interesting developments in numerous directions,
see \eg\ refs. 
[\HohmSamtleben\skipref\BlairMalekPark\skipref\StricklandConstableSubsectors\skipref\BlairMalekRouth\skipref\FreidelLeighMinic-\BermanBlairMalekPerryODD],
 and a
full listing of all the interesting and relevant work is outside the
scope of this brief note and so we refer the reader to the three
reviews for a review of developments and a fuller list of citations of
this rapidly maturing subject.

Hull and Zwiebach [\HullZwiebachCourant], explored the local symmetries of double field
theory infinitesimally and related them to the local symmetries in
ordinary string theory and in the processes uncovered an interesting
Courant algebroid structure. Given the study of the infinitesimal
symmetries the immediate question was what are the finite local
transformations. The question was answered by Hohm and Zwiebach in
ref. [\HohmZwiebachLarge]. Finite transformations have also been
considered in ref. [\Park]. We will further study the finite local
symmetries looking at various global questions.

\section\local{Local Symmetries in Double Field Theory: An
exegesis}Double field theory was brought into being in order to make
manifest the hidden $O(d,d;\mathbb{Z})$ T-duality symmetry of string
theory. It does so by doubling the space and then combining the usual
metric and NS-NS two-form into a single ``generalised metric" that
transforms linearly under a global $O(d,d;\mathbb{R})$ transformation
through conjugation. The action of double field theory is manifestly
invariant under this transformation. The connection to the normal
spacetime is made through the so called {\it{section condition}} or
{\it{strong constraint}} whereby the coordinate dependence of half the
dimensions is removed and hence takes the $2d$ doubled space back to
$d$-dimensions. More formally the section condition restricts the
theory to live on a maximally isotropic subspace which is the normal
spacetime. Different choices of solution to the section condition then
give the duality related theories. The section condition is crucial in
all that follows with most statements only being true using this
condition. 

One is immediately confronted with several questions. 

\item{$\bullet$} First, shouldn't the theory only possess the $O(d,d)$
symmetry when compactified on a torus? 

\item{$\bullet$} Second, how come it is the continuous
$O(d,d;\mathbb{R})$ group and not the restricted arithmetic subgroup
$O(d,d;\mathbb{Z})$? 

\item{$\bullet$} Third, what about the local symmetries of double field theory?

\item{$\bullet$} Finally, a slightly more sophisticated issue. Given
that double field theory should describe the background of a string it
cannot possess any global symmetries that are not the global part of a
local symmetry. That is, in string theory all global symmetries must
be gauged [\Banks]. There are some important caveats to this and
indeed ideally one should rework the proof in ref. [\Banks] for the string
in the doubled background with the action and vertex operators
described in ref. [\Tseytlin]. Nevertheless, let us assume that this lore
should be true also in double field theory. And so how can we have a
global continuous $O(d,d)$ symmetry? 

All the answers to these questions are related. Given that the action
possesses a global continuous symmetry one should simply hope that it
is actually a local symmetry. In fact one can show that this is the
case! The double field theory action is invariant under a ``local
$O(d,d;\mathbb{R})$ transformation'' of the generalised metric. This
transformation is just the usual combination of $d$-dimensional
diffeomorphisms and 2-form gauge transformations (up to section
condition). In double field theory the infinitesimal version of this
transformation is then given by the so called {\it{generalised Lie
derivative}}. Note, it is not the Lie derivative of the generalised
space. And so we see why the {\it{generalised Lie derivative}} is
somewhat of a slight misnomer. It provides an infinitesimal, local,
$O(d,d;\mathbb{R})$ transformation, in the same sense that ordinary
diffeomorphisms induce local $GL(d;\mathbb{R})$ transformations.
To be more concrete, it is not any local $O(d,d)$ transformation but
one constructed from  
the ``usual" tensor transformation matrix,
$M_M{}^N={\*X^N\over\*X'^M}$ and the $O(d,d)$ structure, $\eta$. (The
details of the induced transformation are given in section 3). This
then answers the third question. The local symmetries are the usual
ones but described as a particular realisation of a local $O(d,d)$
transformation. This also 
fits nicely with recent work by Lee and Park on double field theory as
a gauged spacetime [\ParkLeeGauge]. 

In the above discussion there is no mention as to any toroidal
compactification; the $O(d,d)$ is simply a continuous symmetry such
that when made local it can describe the known local symmetries on the
space. This leads to the first two questions on the list. What then if
we do compactify the theory on a torus? As usual we will be left with
the mapping class group. Normally the mapping class group arises from
the global diffeomorphisms not connected to the identity that preserve
the manifold. Here it is the global $O(d,d)$ transformations not
connected to the identity. This then breaks the local
$O(d,d;\mathbb{R})$ to global $O(d,d;\mathbb{Z})$ giving the normal
T-duality group. 

The use of this local $O(d,d)$ symmetry is to construct T-folds [\HullT], exotic
branes 
[\Shigemori\skipref\ShigemoriEB\skipref\Kikuchi-\KikuchiGL,\HohmLustZwiebach]
and other non-geometric backgrounds. To do so we require a
non-contractible one-cycle. We then have non-trivial holonomies of the
generalised metric of the $2d$ doubled space around this one
cycle. This will correspond to an element of
$O(d,d;\mathbb{Z})$. This is normally identified as an element of the
T-duality group. The T-duality group though is a global symmetry and
so it does not really make sense to have a holonomy for a global
symmetry. In fact it is a holonomy in the local $O(d,d)$ symmetry
which due to the topology of the space we can then identify with the
arithmetic T-duality group.  The reduction to the arithmetic subgroup
is due to having a torus fibred over the one cycle. The above
discussion indicates that one could also search for other solutions in
double field theory where the holonomies are other subgroups of
$O(d,d;\mathbb{R})$. Obviously, the fibres must then have different
topology to the torus. The construction of such objects is currently
not known. This maybe the way to realise so called mirrorfolds [\HullT] where
we have a Calabi--Yau space as a fibre that becomes identified with its
mirror upon going round the one cycle of the base, see also  [\Sugawara]. 

And so we have a manifold whose coordinates transform as normal under
general coordinate transformations. The generalised metric though (and
all other generalised tensors) transform under local $O(d,d)$. (The
induced $O(d,d)$ transformation for a given coordinate transformation
while be described later). Is double field theory then some sort of
$O(d,d)$ fibre bundle? The answer is no.  

The doubled manifold is a manifold equipped with an $O(d,d)$
structure. We have a set of coordinate patches $\{ U_\alpha \}$ such
that in each patch the space is isomorphic to $\mathbb{R}^{2d}$ and so
we may introduce coordinates $X_{(\alpha)}^I$, $I=1,\ldots,2d$. (The
signature is dependent on whether one wishes to double time or not;
the subject of which would require a more involved discussion than we
wish to get into in this paper. For now we will just consider doubling
Euclidean space). The $O(d,d)$ structure normally denoted by
$\eta_{IJ}$ then acts as a polarisation and so permits the
coordinates on each patch to be decomposed into pairs so that
$X^I=(x^i,\tilde{x}_i)$ (with $i=1..d$) and where  $X^I \eta_{IJ} X^J=
x^i \delta_i^j \tilde{x}_j $.

On the overlap of any two patches $U_\alpha \cap U_\beta$ we have
transition functions which transform the coordinates from one patch
into the coordinates of the other. These transition functions are
simply diffeomorphisms. That is $X_{(\alpha)}^I(X_{(\beta)})$ is
invertible. As with everything though we only allow diffeomorphisms
that obey the section condition. Since coordinates are not tensors we
will state more explicitly what this means. On an overlap we may
define the difference $\delta
X_{(\alpha\beta)}^I=X_{(\alpha)}^I-X_{(\beta)}^I$. This is a tensor
and it should obey the section condition. The composition of
transition functions on triple overlaps is called the cocycle
condition and for a manifold it must be trivial.  It remains somewhat
mysterious that the transformations in $O(d,d;\mathbb{Z})$ applied
specifically to the overlaps in the geometries named ``genuinely
non-geometric'' in ref. [\HohmLustZwiebach], seem not to obey the
strong section condition, and thus can not be formed as finite
generalised diffeomorphisms, as defined in ref. [\HohmZwiebachLarge]
and the present paper. 

Now, we wish to put a metric on the manifold and so we introduce the
tangent bundle, $F$. As such we proceed as with the usual construction
where we introduce a set of  local trivialisations, denoted by
$\{U_\alpha\}$ which maps the tangent bundle to the local product, of
the manifold, $M^{2d}$ (the base) and its tangent space $TM$ (the
fibre). The presence of the $O(d,d)$ structure on $M$ then also allows
us the decompose the tangent space of the 2d dimensional manifold into
a direct sum of the tangent space and cotangent space of a single $d$
dimensional manifold \ie, $TM^{2d}=TM^d \oplus T^*M^d$. The generalised
metric lives in this space. Subsequently applying the section
condition on $F$ so that we restrict to the $d$-dimensional maximally
isotropic submanifold then leads to so called ``generalised geometry".  
For a more mathematically complete description of the geometry of
double field theory the reader is referred to
refs. [\VaismanI\skipref\VaismanII-\VaismanIII]. 

The bundle is then equipped with transition functions $f_{\alpha
\beta}$ which act on the fibres and so the generalised metric. In
usual geometry these would be the normal diffeomorphisms acting on a
tensor induced from the coordinate transformations between patches
\ie, ${{\partial X_{(\alpha)}^I} \over {\partial X_{(\beta)}^J}} $.  

However, we now have an $O(d,d)$ structure, $\eta$ that must be
preserved; in other words it is globally defined. This means that to
preserve $\eta_{IJ}$ a coordinate transformation must crucially only
induce an $O(d,d)$ transformation on tensors. This is exactly what
happens, under a coordinate transformation in the doubled space, all
tensors transform under a local $O(d,d)$ transformation and so the
transition functions, $f_{\alpha \beta}$ on the fibres now lie in
$O(d,d)$. The precise form of the induced $O(d,d)$ transformation is
given in the following section.

We will show in section 4 that the transition functions between
patches, $ f_{\alpha \beta} \in O(d,d)$  need not obey the cocycle
condition on triple intersections. That is: 
 $$
{\rm{On}} \qquad U_{\alpha} \cap U_{\beta} \cap U_{\gamma}\,: \quad
 f_{\alpha \beta} f_{\beta \gamma} f_{\gamma \alpha}= 1
\eqn
$$ 
may
 {\bf{not}} hold (even up to section condition) as it must in order to
 be a fibre bundle. (The cocycle condition on quadruple intersections
 will be trivial however.) This implies double field theory possesses
 a gerbe structure [\Hitchin] and leads to the following question: If
 it is a gerbe and valued in $O(d,d)$ then is it a non-abelian gerbe
 [\AschieriEtAl]? The answer as we will show is no for a slightly
 subtle reason. The section condition effectively abelianises the
 gerbe in a way described later in this paper. This makes perfect
 sense if one thinks in terms of the ordinary symmetries of the NSNS
 sector. The diffeomorphisms have trivial gerbe structure \ie, obey the
 cocycle condition on triple intersections but the NS two form
 potential is the connection on a local trivialisation for an abelian
 gerbe and its gauge transformations need not. The $O(d,d)$
 transformation on the doubled metric combines these two and so indeed
 one expects there to be a gerbe structure but only an abelian
 part. For a previous discussion on gerbes and their relation to
 T-duality see refs. [\Belov,\HullEtAlGerbes]. 
The presence of the gerbe structure on triple
 overlaps may be connected to the results of Blumenhagen {\it et al.}, L\"ust
 {\it et al.} and Szabo {\it et al.} 
on non-associative deformations in string theory,
 see refs. [\Blumenhagen,\Bakas,\LustNA-\Szabo] and references therein). An
 appropriate mathematical structure for understanding double field
 theory may be that of the Lie algebroid as discussed in
 refs. [\BlumenhagenAlgebroid,\BlumenhagenAlgebroidII]. 

There is a related issue that can further obfuscate matters. Given a
non-trivial coordinate transformation the resulting $O(d,d)$ tensor
transformation may be trivial. In other words, there are coordinate
transformations that are trivial for the generalised metric. One may
see this already at the infinitesimal level with the generalised Lie
derivative
[\BermanCederwallKleinschmidtThompson,\Park,\HohmZwiebachLarge]. It
means there is effectively an equivalence class of coordinate
transformations for each $O(d,d)$ transformation. The consequences were
explored extensively in the seminal work of ref. [\HohmZwiebachLarge] where
the exponential of the generalised Lie derivative was first compared
with a conjectured induced $O(d,d)$ transformation on tensors. We
follow this work closely in the next section and show that one can
identify the exponentiated generalised Lie derivative with the induced
$O(d,d)$ transformation of ref. [\HohmZwiebachLarge] up to an, $O(d,d)$
trivial, coordinate transformation. This is shown to be true to all
orders in the infinitesimal parameter describing the coordinate
transformation. This identification [\HohmZwiebach] was previously
done explicitly
up to quartic order, although a different argument [\Park] has also
been given for its complete consistency.

In the section that follows, we then look at the triple overlap
structure and show how the gerbe structure arises and how the section
condition abelianises the gerbe. 

Finally we end with some comments on the extension of these ideas to
the extended, exceptional geometries that occur in M-theory where the
full U-duality groups are made a manifest symmetry 
[\HullM\skipref\PachecoWaldram\skipref\Hillmann\skipref\BermanPerryGen\skipref\BermanGodazgarPerry\skipref\BermanGodazgarGodazgarPerry\skipref\BermanGodazgarPerryWest\skipref\CoimbraStricklandWaldram-\CoimbraStricklandWaldramII,\BermanCederwallKleinschmidtThompson,\ParkSuh\skipref\CederwallI\skipref\CederwallII\skipref\SambtlebenHohmI-\SambtlebenHohmII].

\bigskip
\bigskip
\bigskip

\section\Limits{The limitations of this analysis}This paper does not
describe how one may patch together the total doubled space to allow a
geometrisation of an arbitrary gerbe. This is a notorious problem and
one that is explored in great detail in the paper [\Papadopoulos]. In
[\Papadopoulos] (which appeared while this paper was in preprint form
on the arXiv) a global obstruction in double field theory is discussed
that would indicate that a construction of the 3rd cohomology classes
associated to three form fluxes requires a highly constrained topology
on the additional dual space. The transformations described in this
paper, for example in section five, (see equation 5.9) are really for
cohomologically trivial fluxes since the gauge transformations of B
described in section five are exact. (In other words we are missing
the transformations corresponding to closed but non exact two forms.)
This along with an absence of discussing the topology in the extended
space limits our analysis.  Such an analysis is beyond this paper and
is an important open question in this area.

Note, even more recently a paper [\Hullmxa] has appeared (several
months after this paper appeared on the arXiv), that analyses some of
these global questions in detail and the proposals we describe here).

\section\TransfSect{Infinitesimal and finite 
generalised diffeomorphisms}Generalised diffeomorphisms are generated
by the generalised Lie derivative, or Dorfman bracket, and are
parametrised by a doubled vector $\xi^M$, encoding diffeomorphisms as
well as $B$-field gauge transformations.
We will adopt the notation of ref. [\HohmZwiebachLarge], where $\xi$
denotes the vector field $\xi^M\*_M$ and where the matrix $a$ is defined
as 
$$
a_M{}^N=\*_M\xi^N\punkt\eqn
$$
The $O(d,d)$ structure allows for a globally defined metric
$\eta_{MN}$, which we may  
use to define a ``transposition" of matrices given by
$$
(a^t)_M{}^N=\eta_{MP}\eta^{NQ}a_Q{}^P\punkt\eqn
$$
In what follows we will always assume that the derivatives satisfy the
strong section 
condition, 
$$
\eta^{MN}\*_M A\*_N B=0\komma\eqn
$$ 
for all $A$ and $B$.

The generalised Lie derivative on a vector $V^M$ is
$$
\LL_\xi V^M=\xi V^M-V^N(a-a^t)_N{}^M\komma\eqn
$$ and for a covector $W_M$
$$
\LL_\xi W_M=\xi W_M+(a-a^t)_M{}^NW_N\punkt\eqn
$$
It will be useful to write this in a compact form as follows,
$$
\LL_\xi=\xi+a-a^t=L_\xi-a^t\komma\Eqn\DoubleLieDerEq
$$
where $L_\xi$ is the ordinary Lie derivative. The generalised Lie
derivatives are derivatives in that they obey the Leibniz rule for
products. The generalised Lie derivative on a scalar is the same as
the Lie derivative, \ie, it is simply the so called translation term,
$\xi$ acting on the scalar. (Note that the generalised Lie derivative
on $\eta$ vanishes).

The algebra of generalised diffeomorphisms is (up to section condition)
$$
[\LL_\xi,\LL_\chi]=\LL_{\leftbr\xi,\chi\rightbr}\komma\eqn
$$
with $\leftbr\xi,\chi\rightbr=\fr2(\LL_\xi\chi-\LL_\chi\xi)$.

A short calculation shows that we can equally
well express the 
commutator in terms of the ordinary Lie bracket of vector
fields (Lie derivatives) together with an extra term, $\Delta$. We then have
$$
[\LL_\xi,\LL_\chi]=\LL_{[\xi,\chi]}+\Delta_{\xi,\chi}\komma
\Eqn\AltComm
$$
where $\Delta_{\xi,\chi}$ has the simple form
$$
\Delta_{\xi,\chi}=-ab^t+ba^t\komma\Eqn\FormOfDelta
$$
with $a_M{}^N=\*_M\xi^N$, $b_M{}^N=\*_M\chi^N$.

Since $\LL_{[\xi,\chi]}$ in the right hand side of eq. (\AltComm) is a
gauge transformation, so is $\Delta$. $\Delta$ represents a generalised
coordinate transformation with parameter
$\zeta^M=-\xi^N(b^t)_N{}^M$. Crucially the translation term vanishes
for $\Delta$. Thus it has the important property that it affects
neither coordinates nor derivatives. This is because 
$$
a^t\*=0\eqn
$$
thanks to the section condition. 
Note that, the multiplication of any two matrices of the form (\FormOfDelta)
gives zero. Eq. (\AltComm) is the infinitesimal version of what we
will later, for finite transformations, associate to a gerbe.
Note that exponentiation of these non-translating transformations is
simple, the nilpotent property of $\Delta$ immediately gives 
$$
e^\Delta=1+\Delta\punkt\Eqn\ExpDeltaEq
$$

Hohm and Zwiebach [\HohmZwiebachLarge] 
conjectured an explicit expression for
the finite transformation of tensors under a generalised
diffeomorphism in doubled geometry. When $X\rightarrow X'(X)$, a
covector transforms as 
$$
W_M'(X')=F_M{}^N(X',X)W_M(X)\komma\eqn
$$
where
$$
F_M{}^N(X',X)=\fr2\left(M\ti M+\ti M M\right)\komma\Eqn\FDef
$$
$M_M{}^N$ being the ``usual'' transformation matrix, 
$$
M_M{}^N={\*X^N\over\*X'^M}\punkt\eqn
$$
The section condition ensures, quite non-trivially, that the
conjectured transformation matrix $F$ is a group element of $O(d,d)$. 
We will now demonstrate that the expression (\FDef) provides the
correct expression for the exponentiation of the generalised Lie
derivative. More precisely, we will show that the finite
transformation obtained from the exponentiated Lie derivative lies in
the same equivalence class as $F$ modulo non-translating transformations.

When working with expressions in terms of the matrices $a$ and $a^t$,
it is important to recall that they satisfy $a^ta=0$ due to the section
condition. One can therefore replace any matrix  that is produced of
$a$ and $a^t$ by its ordered expression with all the $a$'s to the left
and all the $a^t$'s to the right, in a way reminiscent of normal
ordering of operators in a quantum mechanics. Any matrix which is
obtained only from matrix multiplication of the basic ingredients $a$,
$a^t$ can be put on such a form. We may identify the form of a matrix 
$f(a,a^t)$
given by this ordering prescription with a ``symbol'' $S[f](x,y)$ 
of the matrix. Here, $x$ and $y$ are not matrices but commuting formal
variables. 
This identification of course yields $S[a]=x$, $S[a^t]=y$, and \eg\ 
$S[aa^t]=xy$, $S[a^ta]=S[0]=0$.
The matrix product is reproduced by an associative star product of symbols:
$$
S[fg]=S[f]\star S[g]\komma\eqn
$$
where
$$
(A\star B)(x,y)=A(x,y)B(0,y)+A(x,0)B(x,y)-A(x,0)B(0,y)
\komma
\Eqn\OrderedProdEq
$$
allowing for index-free calculations.

We can as a warm up exercise exponentiate the matrix part
of the generalised Lie derivative,
$$
e^{a-a^t}=\sum\limits_{n=0}^\infty\sum\limits_{i=0}^n
       {1\over n!}a^{n-i}(-a^t)^i\komma\eqn
$$
so we find its symbol to be
$$
\eqalign{
S[e^{a-a^t}](x,y)&={1\over x+y}\sum\limits_{n=0}^\infty{1\over n!}
       \left(x^{n+1}-(-y)^{n+1}\right)\cr
&={xe^x+ye^{-y}\over x+y}\komma\cr
}\Eqn\GExpr
$$
The seemingly singular behaviour of the denominator is of course
compensated by the nominator.

The transformation matrix $M$ can be expressed as
$$
M=e^{-\xi}e^{\xi+a}\komma\Eqn\MExprEq
$$
where the first factor translates back to the original coordinate.
In these expressions, the operator $\xi$ acts to everything on the
right. Some care has to be taken when transposing an operator. In what
follows we will also need the transpose of $M$, which is 
$$
M^t=e^{-\xi+a^t}e^\xi\punkt\eqn
$$
The analogous expression to $M$ for the generalised Lie derivative is,
$$
G=e^{-\xi}e^{\xi+a-a^t}\punkt\eqn
$$
This is what we now want to compare to $F$ in eq. (\FDef). Using the
section condition, we may rewrite the matrix $F$ in the normal ordered
form as follows: 
$$
\eqalign{
F&=\fr2\left(M(M^{-1})^t+M+(M^{-1})^t-1\right)\cr
&=\fr2\left(e^{-\xi}e^{\xi+a}e^{-\xi}e^{\xi-a^t}
   +e^{-\xi}e^{\xi+a}+e^{-\xi}e^{\xi-a^t}-1\right)\punkt\cr
}\eqn
$$
(Here, and in the following,  we have chosen not to use the symbols introduced earlier in
this section, but use explicit $a$'s and $a^t$'s. When needed,
subscripts $R$ and $L$ are used to denote right and left multiplication.)
To compare to $F$, we need to put the matrix $G$ also in normal ordered form. 
We therefore write 
$$
e^{\xi+a-a^t}=\sum\limits_{n=0}^\infty{1\over n!}(\xi+a-a^t)^n\komma\eqn        
$$
and use
$$
\eqalign{
(\xi+a-a^t)^n&=(\xi+a)^n+\sum\limits_{i=1}^n(\xi+a)^{n-i}(-a^t)(\xi-a^t)^i\cr
&=(\xi+a)^n-{(\xi+a)_L^n-(\xi-a^t)_R^n\over(\xi+a)_L-(\xi-a^t)_R}\,a^t\cr
&={(\xi+a)^na+a^t(\xi-a^t)^n\over(\xi+a)_L-(\xi-a^t)_R}\komma\cr
}
\eqn
$$
where the first step is easily shown by induction, and where $R$ and
$L$ denote that the operators stand on the far right and left. 
Therefore, the exponentiated
transformation takes the form
$$
e^{\xi+a-a^t}={e^{\xi+a}a+a^te^{\xi-a^t}\over(\xi+a)_L-(\xi-a^t)_R}\punkt\eqn
$$

We now use the expression (\OrderedProdEq) 
for products of ``functions'' of $a$ and
$a^t$. The operator $\xi$ is a scalar in the sense that it commutes
with $a^t$, after using the section condition, and so
eq. (\OrderedProdEq) for the product of 
ordered expressions still holds. Then after some calculations one find
$$
G^{-1}F=1+\fr2{e^{-(\xi+a)}(a-a^t)e^{\xi-a^t}-(a-a^t)\over(\xi+a)_L-(\xi-a^t)_R}
+\fr2(e^{-\xi}e^{\xi-a^t}-e^{-(\xi+a)}e^\xi)
\punkt\Eqn\GInverseF
$$
Despite the appearance of negative powers of operators, the series
expansion contains only positive powers and is well defined.
It is not yet obvious to why this is of the form we want it to be,
namely a transformation which is a product of factors of the form
(\ExpDeltaEq).
Note that the second term in eq. (\GInverseF) can be written on
the form $f(\ad_\xi+a_L-a^t_R)\cdot(a-a^t)$, where
$f(z)=z^{-1}(e^{-z}-1)$, and that the r\^ole of the last term is to
remove from this the parts that contain only $a$ and only $a^t$ (\ie,
the parts obtained by formally setting $a^t$ or $a$ to zero). These
properties will be essential for the proof.

In order to better understand what this implies, we investigate the
expression $(\adxi+a_L+a^t_R)^n\cdot a$ and its transpose
(here, the dot denotes the operator acting only on the following
matrix/operator, not on everything on the right), occurring
in an expansion of the function $f$. For any operator $b$, it is
straightforward to show by induction that
$$
(\adxi+a_L+a^t_R)^n\cdot b=(\adxi+a)^n\cdot b
+\sum\limits_{k=0}^{n-1}{n\choose k}((\adxi+a)^k\cdot b)
     ((\adxi+a^t)^{n-k-1}\cdot a^t)\komma\eqn
$$
so eq. (\GInverseF) may be written as
$$
\eqalign{
&G^{-1}F=1+\fr2\sum\limits_{n=1}^{\infty}
\sum\limits_{k=0}^{n-1}{(-1)^{n+1}\over(n+1)!}{n\choose k}
  ((\adxi+a)^k\cdot a)((\adxi+a)^{n-k-1}\cdot a)^t-(\ldots)^t\cr
&=1+\fr2\sum\limits_{n=2}^{\infty}
\sum\limits_{k=0}^{n-1}{(-1)^n(n-2k-1)\over(n+1)(k+1)!(n-k)!}
  ((\adxi+a)^k\cdot a)((\adxi+a)^{n-k-1}\cdot a)^t\punkt\cr
}
\eqn
$$
Finally, we need to reinterpret the factors
$(\adxi+a)^k\cdot a$, in order to ensure that the terms are of the
desired form (\FormOfDelta), with both factors being derivatives of
vectors. 
Explicit expansion yields unwieldy expressions, of which
the lowest ones are
$$
\eqalign{
(\adxi+a)^0\cdot a&=a\komma\cr
(\adxi+a)^1\cdot a&=a^2+[\xi,a]\komma\cr
(\adxi+a)^2\cdot a&=a^3+[\xi,a]a+2a[\xi,a]+[\xi,[\xi,a]]\komma\cr
(\adxi+a)^3\cdot a&=a^4+[\xi,a]a^2+2a[\xi,a]a+3a^2[\xi,a]\cr
     &+[\xi,[\xi,a]]a+3[\xi,a][\xi,a]+3a[\xi,[\xi,a]]
     +[\xi,[\xi,[\xi,a]]]   \punkt\cr
}\Eqn\XiXiXi
$$
Inspection of the first of these expressions leads to the guess
$$
(\adxi+a)^na=\*(\xi^n\cdot\vxi) \, , \Eqn\AdXiN 
$$ 
where the vector notation denotes $(\*\vvecc v)_M{}^N=\*_Mv^N$.
It is straightforward to then prove this by induction. This implies that
$G^{-1}F$ is the product of matrices of the form
(\ExpDeltaEq), and thus is a finite non-translating generalised
diffeomorphism connected to the identity. Explicitly, 
$$
G^{-1}F=\prod\limits_{n=2}^{\infty}
\prod\limits_{k=0}^{n-1}\left(1+\fr2{(-1)^n(n-2k-1)\over(n+1)(k+1)!(n-k)!}
  \*(\xi^k\cdot\vxi)(\*(\xi^{n-k-1}\cdot\vxi))^t\right)
\punkt
\eqn
$$
Observe that the sum (\GInverseF) may be replaced by a product as has
been done above since the product of any two terms vanishes via the
section condition. 
Each term in $G^{-1}F$ is of the form given in (\FormOfDelta) and
therefore does not involve a translation. Thus $G^{-1}F$ is in the
same equivalence class of transformations as the identity. 
 
In order to compare to previous results it is useful to explicitly
evaluate this expression as a series expansion. To do this, first the
expression f or $G^{-1}F$ is written as a
function of two variables: Let $x$ represent the first $\xi$ and $y$
the second. Then the deviation from unity is encoded in the function:
$$
\eqalign{
T(x,y)&-1={(e^{-x}+1)(e^{-y}-1)x-(e^{-x}-1)(e^{-y}+1)y\over2(x+y)}\cr
&=-\fr{12}(x^2y-xy^2)+\fr{24}(x^3y-xy^3)-\fr{80}(x^4y-xy^4)
-\fr{120}(x^3y^2-x^2y^3)\cr
&\qquad+\fr{360}(x^5y-xy^5)+\fr{288}(x^4y^2-x^2y^4)\cr
&\qquad-\fr{2016}(x^6y-xy^6)-\fr{1120}(x^5y^2-x^2y^5)-\fr{2016}(x^4y^3-x^3y^4)
+\ldots\cr
}\eqn
$$

A series expansion of eq. (\GInverseF) to third order in the
parameter confirms the earlier results of ref. [\HohmZwiebachLarge]:
$$
G^{-1}F=1-\fr{12}(a^2a^t+[\xi,a]a^t-(\ldots)^t)
+O(\xi^4)\eqn
$$
This is a non-translating transformation with parameter
$$
\zeta_3^M=-\fr{12}\xi^N(aa^t)_N{}^M\komma\Eqn\ZetaThree
$$
which relies on the appearance of the combination
$a^2a^t+[\xi,a]a^t$. The above demonstrates that this happens to all
orders for finite transformations connected to the identity.

\section\OverlapSect{Overlaps and gerbes}The matrix $F$
may be more clearly associated to a specific finite change
of coordinates by writing it $F(M)$. It is a group element in
$O(d,d)$, and $F(M)F(M^{-1})=1$.  

Composition of transformations is less straightforward. Remember that
$M={\*X'\over\*X}$. From the transformation of the coordinates one
might expect the chain rule to hold, 
$$
{\*X''\over\*X}={\*X'\over\*X}{\*X''\over\*X'}\komma\Eqn\Chainrule
$$
but $F$ generically fails to
respect the expected corresponding composition rule, \ie, 
$$
F(M)F(N)\neq F(MN)\, . \Eqn\CompositionFailureEq
$$
That is, the map $F:\,GL(2d)\rightarrow O(d,d)$ is not a group homomorphism.
This has led to attempts [\HohmZwiebachLarge,\HohmLustZwiebach] 
to modify the product of two $F$'s in terms of a
``$*$-product'', which turns out to be non-associative.

The way we would rather understand composition, is that $F(M)$ is one
representative in a class of transformations, all transforming the
coordinates identically, but differing by non-translating
transformations as in the previous section. Composition of
transformations $F(M)$ and $F(N)$ 
will result in a third transformation which is in the
equivalence class dictated by the composition (\Chainrule). It will
however not be the representative given by $F(MN)$, but differ from it
by some non-translating transformation of the special type discussed
in the previous section. In other words, the composition rule holds up
to a non-translating transformation. Schematically,
eq. (\CompositionFailureEq) 
is replaced by a ``twisting'' by a non-translating transformation:
$$
F(M)F(N)=F(MN)\hbox{exp}\bigl(\sum_i\Delta_i\bigr)\punkt\eqn
$$
Composition of equivalence classes is obeyed, but we need to make the
detailed structure clearer.
The behaviour points towards a description in terms of gerbes, where
$e^\Delta$ represents the cocycle.

Let us think of the matrix $F(M)$ as providing the transformation
relating fields (\eg\ a generalised metric) on the intersection of two
patches. In order to isolate the deviation from the chain rule, 
we consider instead the overlap as defined by\foot{This is in
itself a triple overlap, with one patch fixed [\AschieriEtAl]. 
The preference of this
form over $F$ lies in the fact that the matrices around $F(M^{-1}N)$
compensate for the indices belonging to different frames.}
$$
H(M,N)=F(M)F(M^{-1}N)F(N^{-1})\punkt\Eqn\HDef
$$
In standard language, this would be identified with an overlap
$f_{\a\b}$ between patches $\a$ and $\b$. Clearly, eq. (\HDef)
satisfies
$$
H(M,N)H(N,M)=1\komma\eqn
$$
corresponding to $f_{\a\b}f_{\b\a}=1$. 

A triple overlap is locally written as
$\l_{\a\b\g}=f_{\a\b}f_{\b\g}f_{\g\a}$.
This amounts to forming
$$
\eqalign{
\Lambda(M,N,P)&=H(M,N)H(N,P)H(P,M)\cr
&=F(M)F(M^{-1}N)F(N^{-1}P)F(P^{-1}M)F(M^{-1})\komma\cr
}\eqn
$$
We want to evaluate this expression, using the basic form (\FDef) of
$F$. This is straightforward but somewhat tedious work. The section
condition will, as usual, be necessary, and tricks similar to
eq. (\OrderedProdEq) are helpful. The matrix $H(M,N)$ turns 
out to have the form
$$
H(M,N)=1-\fr2(mn^t-nm^t)=1+Q(M,N)\komma\eqn
$$
where $M=1+m$, $N=1+n$.
This is a group element of $O(d,d)$ of a special form, $1$ plus an
antisymmetric matrix $Q$, which is nilpotent, $Q^2=0$, thanks to the
section condition, leading to $HH^t=(1+Q)(1-Q)=1-Q^2=1$.
Furthermore, in the multiplication of any two $H$'s, their ``$Q$
terms'' will simply add, as the product of the $Q$'s is zero.
The matrix takes 
the same form as the ones encountered earlier in eq. (\ExpDeltaEq).
Therefore, the rest of the calculation leading to the triple overlap
$\Lambda$ is simple. We find
$$
\Lambda(M,N,P)=1-\fr2(mn^t+np^t+pm^t-nm^t-pn^t-mp^t)\punkt\eqn
$$  
Now $\Lambda$ behaves as an overlap should, with manifest properties under
permutation. In a certain sense, the matrices $H$, although being
group elements of $O(d,d)$, behave in an abelian way. So a cocycle
condition on $\Lambda$ can be written down without any problem. This is
slightly problematic had the gerbe turned out to be non-abelian 
[\AschieriEtAl]. Let
us investigate this gerbe structure by explicitly picking a solution
to the section condition. 

Divide the index $M$ into two groups of light-like ones,
$(m,\bar m)$, with $\eta_{m\bar n}=\delta_{mn}$. Then solve the
section condition by 
letting the derivative carry unbarred indices only.
The form of the matrix $H$ is then
$$
H_M{}^N=\left(\matrix{H_m{}^n&H_m{}^{\bar n}\cr
                      H_{\bar m}{}^n&H_{\bar m}{}^{\bar n}}\right)
=\left(\matrix{\delta_m{}^n&-\fr2(\*_m\xi^P\*^{\bar n}\chi_P
                                 -\*_m\chi^P\*^{\bar n}\xi_P)\cr
               0&\delta_{\bar m}{}^{\bar n}}\right)\punkt\eqn
$$
Here it is even more
explicit that the non-trivial terms on the off-diagonal are additive,
and that they vanish for transformations involving $x^m$ only. This
shows that the gerbe structure is confined to gauge transformations of
the $B$-field. This can be seen explicitly by acting with $H$ on the
generalised metric through conjugation and seeing that it transforms
$B$-field. The gerbe structure connected to $B$-field gauge
transformation was obtained in generalised (not doubled) K\"ahler
geometry in  
ref. [\HullEtAlGerbes].

This shows how a doubled manifold, before it is equipped with a
generalised metric structure, is an ordinary manifold (subject to some
additional constraints due to the presence of a global $O(d,d)$
structure), and may be 
patched together with ordinary overlaps. Once a solution to the
section condition is specified, the topology in fact becomes that of
a fibre bundle. Introduction of a generalised metric (or other
tensors) refines the structure and introduces a gerbe over the doubled
manifold. Earlier discussion on non-associative structures are a
reflection of the presence of a non-trivial triple overlap.

\section\ExceptionalSect{Some comments on exceptional/extended
geometry}It 
is desirable to perform the corresponding investigation
also for the exceptional/extended geometry described in
refs. [\HullM\skipref\PachecoWaldram\skipref\Hillmann\skipref\BermanPerryGen\skipref\BermanGodazgarPerry\skipref\BermanGodazgarGodazgarPerry\skipref\BermanGodazgarPerryWest\skipref\CoimbraStricklandWaldram-\CoimbraStricklandWaldramII,\BermanCederwallKleinschmidtThompson,\ParkSuh\skipref\CederwallI\skipref\CederwallII\skipref\SambtlebenHohmI-\SambtlebenHohmII]. We
refer the reader to those papers or the reviews
[\Marques,\BermanThompson] for an introduction. In what follows we will
discern some similarities but also some 
differences when trying to generalise the finite transformations for
$O(d,d)$ to the exceptional groups associated with U-duality. This is
ongoing work and the final finite form of the induced transformation
has not been yet been determined but it is worthwhile showing how some
similar structures arise. 

In order to make the formalism as close to the one employed in doubled
geometry as possible, we note that the generalised Lie derivative
takes the form
$$
\LL_\xi=\xi+a-a^Y=L_\xi-a^Y\komma\eqn
$$
which looks very similar to eq. (\DoubleLieDerEq), but where the
transpose is replaced by the operation 
$$
a\rightarrow a^Y\,,\quad (a^Y)_M{}^N=Y_{MP}{}^{QN}a_Q{}^P\komma\eqn
$$ 
$Y$ being the $E_{n(n)}$-invariant tensor (in double geometry, 
$Y_{MN}{}^{PQ}=\eta_{MN}\eta^{PQ}$) responsible for the section
condition,
$Y_{MN}{}^{PQ}\*_P\ldots\*_Q=0$; and so again we will require:
$$Y_{MN}{}^{PQ}\*_P A \*_Q B=0$$ for all $A$ and $B$. The $Y$-tensor
also fulfills some algebraic 
relations, which ensure the closure of the algebra of generalised Lie
derivatives (see ref. [\BermanCederwallKleinschmidtThompson]). 

One difference, which seems to lie behind most difficulties in writing
a closed form for the finite transformations, is that, 
unlike the transpose in $O(d,d)$, the operation $a\rightarrow a^Y$ is not an
involution. It certainly has eigenvalue $1$ on the part of a matrix
outside the adjoint of $E_{n(n)}\times\RR^+$, so that $a-a^Y$ is an
element in the Lie algebra, but other eigenvalues are different from
$-1$.

This has so-far prevented us from finding a closed candidate for the
transformation $F$, corresponding to eq. (\FDef). 
Some considerations can be done at the level of infinitesimal 
transformations however.
If we construct a non-translating transformation from a parameter
$\vzeta=\vxi b^Y=\vxi(\*\veta)^Y$, we get a Lie algebra element
$$
\*\vzeta-(\*\vzeta)^Y=ab^Y-(ab^Y)^Y\punkt\Eqn\ExceptionalParam
$$
(It is important to have a derivative of a vector on the right,
otherwise the terms where the derivative acts on that matrix will not
go away. This may be the most general form of a non-translating transformation.)
This is not the same as $ab^Y-ba^Y$. 
It does not seem to be true that such Lie algebra
elements are in general (quadratically) nilpotent. 
This may lead to a higher gerbe structure.

Quite non-trivially, an expression for the algebra of generalised Lie
derivatives in terms of the usual Lie bracket and a remainder term
$\Delta$ may be obtained following a similar calculation leading to
eq. (\AltComm). After quite some work one can find that: 
$$
[\LL_\xi,\LL_\eta]=\LL_{\leftbr\xi,\eta\rightbr}
=\LL_{[\xi,\eta]}+\Delta_{\xi,\eta}\komma\eqn
$$
where the remainder piece $\Delta$ is now given by
$$
\eqalign{
\Delta_{\xi,\eta}&=[a,b]^Y+[a^Y,b^Y]-[a,b^Y]-[a^Y,b]\cr
&=-\fr2\left([a,b^Y]+[a^Y,b]-[a,b^Y]^Y-[a^Y,b]^Y\right)\cr
&=-\fr2\left(ab^Y-ba^Y-(ab^Y-ba^Y)^Y\right)\punkt\cr
}\eqn
$$
This is a useful step in generalising the results from the $O(d,d)$
case as it encouragingly shows an analogous structure exists in the
algebra of generalised Lie derivatives for the exceptional groups. 
In the above, we have used the ``general'' identity
$$
[a,b]^Y+[a^Y,b^Y]-\fr2\left([a,b^Y]+[a,
b^Y]^Y+[a^Y,b]+[a^Y,b]^Y\right)=0
\eqn
$$
where ``general" implies it holds even if the section condition is not
applied. This is a rewriting in terms of matrices of 
identities for the $Y$ tensor of ref. [\BermanCederwallKleinschmidtThompson].

Note, eq. (\ExceptionalParam) is automatically antisymmetric in $a$ and
$b$. This is because of the general identity $[m,m^Y]-[m,m^Y]^Y=0$. It
is easily understood, since $[m,m^Y]$ must be proportional to the
commutator between the adjoint projection of $m$ and its complement,
which is a module, and therefore the result has vanishing adjoint
component. Letting $m\rightarrow a+b$, where the first indices fulfill
the section condition, shows the antisymmetry.

How can a transformation of the type (\ExceptionalParam) be exponentiated
to a finite transformation? What is the final finite form? What are
the global gerbe properties? In fact we do expect the presence of
higher gerbe structures. The 3-form potential $C_{(3)}$ is naturally
the connection on a 3-gerbe which may have a non-trivial 4-overlap. For
duality groups beyond $SL(5)$ the 6-form potential $C_{(6)}$ will play
a role and it comes with a 6-gerbe structure and may have non-trivial
7-overlap. Some preliminary investigation points in this direction. It
remains a key challenge to exceptional/extended geometry to find the
finite form of its local symmetries and understand its global
properties.

\acknowledgements We would like to thank Daniel C. Thompson, Neil
Copland and Edward Musaev for discussions and collaboration at an
early stage. We are also grateful to the following people for general
discussions on double field theory that inspired this paper: Ralph
Blumenhagen, Diego Marqu\'es, Jeong-Hyuck Park, Peter West and Edward
Witten. DSB is partially supported by STFC consolidated grant
ST/J000469/1 and is also grateful for DAMTP in Cambridge for
continuous hospitality.

\refout

\end